# Detection of the Cherenkov light diffused by Sea Water with the ULTRA Experiment

*The ULTRA Collaboration:*
G.Agnetta [1], P. Assis [2], B. Biondo [1,3], P. Brogueira [2], A. Cappa [4,5], O. Catalano [1,3], J. Chauvin [6],
G. D'Alì Staiti [1,3,7], M. Dattoli [4,5,8], M.C. Espirito-Santo [2], L. Fava [5,8], P. Galeotti [5,8], S. Giarrusso [1,3],
G. Gugliotta [1,3], G. La Rosa[1,3], D. Lebrun [6], M.C. Maccarone [1,3], A. Mangano [1,3], L. Melo [2],
S. Moreggia [6], M. Pimenta [2], F. Russo[1], O. Saavedra [5,8], A. Segreto [1,3], J.C. Silva[2], P. Stassi[6],
B. Tomé [2], P. Vallania [4,5], C. Vigorito [5,8]

1. INAF-IASF, Palermo, Italy
2. LIP, Lisbon, Portugal
3. INFN, Catania, Italy
4. INAF-IFSI, Torino, Italy
5. INFN, Torino, Italy
6. LPSC, Grenoble, France
7. DiFTeR, University of Palermo, Italy
8. University of Torino and INFN, Torino, Italy

## Abstract

The study of Ultra High Energy Cosmic Rays represents one of the most challenging topic in the Cosmic Rays and in the Astroparticle Physics fields. The interaction of primary particles with atmospheric nuclei produces a huge Extensive Air Shower together with isotropic emission of UV fluorescence light and highly directional Cherenkov photons, that are reflected/diffused isotropically by the impact on the Earth's surface or on high optical depth clouds. For space-based observations, detecting the reflected Cherenkov signal in a delayed coincidence with the fluorescence light improves the accuracy of the shower reconstruction in space and in particular the measurement of the shower maximum, giving a strong signature for discriminating hadrons and neutrinos, and helping to estimate the primary chemical composition. Since the Earth's surface is mostly covered by water, the ULTRA (UV Light Transmission and Reflection in the Atmosphere) experiment has been designed to provide the diffusing properties of sea water, overcoming the lack of information in this specific field. A small EAS array, made up of 5 particle detectors, and an UV optical device, have been coupled to detect in coincidence both electromagnetic and UV components. The detector was in operation from May to December, 2005, in a small private harbour in Capo Granitola (Italy); the results of these measurements in terms of diffusion coefficient and threshold energy are presented here.

## Introduction

The EUSO [1] experiment has been designed to study the UHE tail of the cosmic ray spectrum, using the avantgarde technique of the Space detection. The cue was given by the contradictory results obtained by the HiRes [2] and AGASA [3] experiments: the lack of the expected GZK cutoff and the event clustering seen by the latter could be the marking of new physics together with the possibility of identifying UHE cosmic ray sources from charged particles. During the Phase A study, approved by ESA and started on March 2002, the EUSO Collaboration realized that the experiment could also measure the UHE neutrinos (that can be used as a probe for cosmological models), and study short term variation atmospheric phenomena recently discovered and still widely unknown (sprites, blue jets and elves).

Besides the EUSO experiment, the AUGER Extensive Air Shower (EAS) array has been approved and almost completed. The first results published by the AUGER Collaboration don't show any evidence of GZK violation or event clustering [4]. Anyway the innovation of the EUSO technique and its scientific goals, only partly coincident with the AUGER ones, implies that only a Space-based experiment will have the chance to study with high statistical significance the expected recovery spectrum beyond the GZK cut-off energy and the UHE source distribution. Its realization appears even more worth to be pursued. After the Phase A completion, the mission was approved by ESA for the scientific case, but it was put in a "frozen" state mainly due to lack of funding. Recently the same project, with minor modifications, has been accepted by the JAXA and NASA Agencies with the JEM-EUSO acronym.

The advantages of the UHE cosmic ray detection from Space are the huge monitored target mass (~10$^{13}$ tons)

and the possibility of observing the fluorescence tracks at equal distance, avoiding the proximity effect typical of ground-based detectors; the inconveniences are the signal dilution due to the distance, that increases the minimum detectable energy, and the high cost. The basic parameters of the EUSO experiment, as of any other EAS detector, are the sensitivity and the energy threshold, depending on the background in the relevant wavelength window and on the fluorescence light attenuation. These effects have been studied experimentally by means of dedicated tests [5]; the goal of the ULTRA experiment is to verify the possibility of detecting the Cherenkov light reflected/diffused (r/d) by the sea surface, giving the possibility of measuring the maximum shower depth and increasing significantly the refinement of the EAS parameters.

## The Detector

A paper describing in detail the detector layout and its performances has been published in [6]; we remind here briefly that the aim of the ULTRA experiment is to measure the r/d coefficient of Cherenkov light by sea water correlating the information coming from 3 different devices:
- a small EAS array to measure the shower size and the arrival direction;
- a wide field Cherenkov detector, pointing to zenith from the centre of the EAS array, to measure the incoming Cherenkov light;
- a narrow field Cherenkov Ultraviolet (UV) detector, pointing to the centre of the EAS array from a higher location, to measure the r/d signal.

The location of the diffused Cherenkov light detector was chosen outside the EAS array to reduce the amount of particles crossing the photomultiplier, and to add a time delay with respect to the true diffused light signal long enough to be discriminated.

The experiment was installed in the private harbour of Capo Granitola (Italy); previous measurements were done in Mont-Cenis (France), to calibrate the EAS array, and in Grenoble (France), to calibrate the Cherenkov detectors by means of a highly diffusing layer of Tyvek 1025D [7].

## Data analysis

The expected signal in the narrow field Cherenkov detector is calculated assuming an isotropic distribution of the r/d signal by the sea surface and a radial symmetry for the produced air shower Cherenkov light (see [8] for further details). Due to the low elevation of the UV telescope with respect to the sea surface, the field of view (FoV) is a very elongated ellipse, and such a geometry of the detection system modifies the expected signal. The Cherenkov light, concentrated in a narrow pulse with less than 10 ns duration, will produce a much broader signal in the detector since the photons hitting the sea surface will be r/d with different time of flight depending on their distance from the UV telescope and on the arrival direction of the shower. The knowledge of this latter parameter, given by the EAS array together with the shower size and core location, allows the measurement of the Reflection Coefficient ($R_C$). $R_C$ is defined as the ratio between $N_{det}$ and $N_{exp}$, where $N_{det}$ is the number of photons detected by the UV telescope, and $N_{exp}$ is the number of expected photons in case of total r/d. $N_{det}$ is obtained from the data knowing the photoelectron conversion and the efficiency of the various elements of the telescope: Fresnel lens, BG3 filter (if present) and PMT quantum and collecting efficiency. All these terms have been separately measured or taken from the specifications. The expected signal is observed dividing the elliptical FoV in narrow slices at equal distance from the detector and adding up all these contributions:

$$N_{\exp} = T_r \cdot T_m \cdot A_{lens} \cdot \sum_i \frac{N_i}{2 \cdot \pi \cdot d_i^2}$$

where $A_{lens}$ is the lens collecting area, $N_i$ is the number of photons inside the i-th slice of the FoV and $d_i$ is the mean distance of the i-th slice from the lens entrance pupil. Due to the small distance between the Cherenkov detector and its FoV, ( ~≤100 m), the Rayleigh ($T_r$) and Mie ($T_m$) scattering have been neglected.

The number of photons falling within the FoV have been evaluated using a Monte Carlo simulation [9]. A total of 50 showers with full electromagnetic and Cherenkov light components were simulated with energy ranging from $10^{15}$ and $5 \times 10^{16}$ eV and zenith angle from 0° to 30°. From these data, a parametrization of the Cherenkov light lateral distribution as a function of the shower size and arrival direction has been obtained. The time profile of the light pulse detected by the Cherenkov telescope is also used to check the data consistency. It depends on both the distance of the detector with respect to the photons position inside the FoV and on the EAS arrival direction. Using the shower size and arrival direction measured by the EAS array, we can obtain the expected pulse shapes and relative timing between the different detectors, on an event by event basis, validating or rejecting the single event. The wide field Cherenkov detector pointing to zenith was thought to give a further validation on the adopted conversion from shower size to Cherenkov photons, but due to the limited pupil entrance its use was hampered by the fluctuations, and for this analysis it was used only for the timing selection.

## Results

Data have been collected from May, 9th to November, 5th, 2005, for a total of 199.5 hours with the BG3 filter and 73.7 hours without it. The small efficiency is due to the constraint of acquiring data only during clear moonless nights. Due to the array dimensions and to the energy threshold of the Cherenkov detectors, the total number of coincident events is 41, reduced to 36 after the timing cuts briefly described in the previous section. Figure 1 shows the core positions of these events, compared with the Cherenkov telescope FoV.

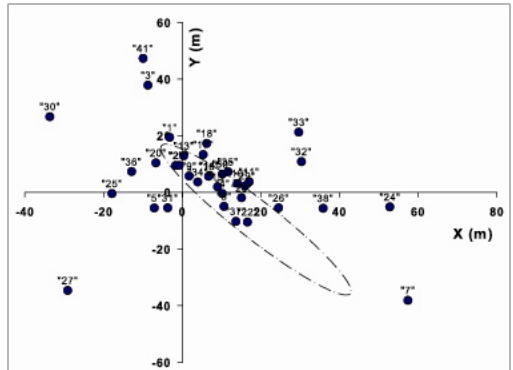

Figure 1: Core location of the coincident events and Cherenkov telescope FoV.

As expected, most of the events have the core located inside or close to the elliptical FoV. However, since some events are far from it, we checked using the measured shower size that these are also the most energetic ones. Figure 2 shows the comparison between the shower size spectrum for all the events with the core located inside the FoV and the fraction of 36 events in coincidence.

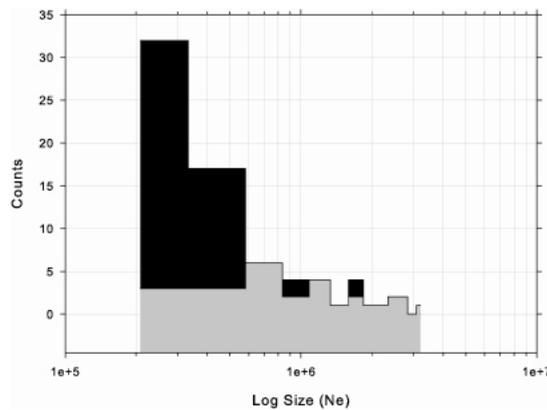

Figure 2: Shower size spectrum for all the events within the UV telescope FoV (black) and for the coincident sub-sample (gray).

The detection efficiency of the r/d Cherenkov light increases with size, reaching full efficiency from $5 \times 10^5$ particles (corresponding to $\sim 8 \times 10^{15}$ eV). Finally, the distribution of the r/d coefficient $R_C$ is shown in Fig.3.
On average, a 4.1 % (with RMS 2.2%) of light is r/d from the sea surface. No correlation of the $R_C$ coefficient with the core location, the arrival direction or the shower size is found. However, the high reflection coefficient tail is due to events with the core very close to the centre of the FoV, possible connected to the core shower development inside the water. Unfortunately the very low statistics (3-5 events) does not allow any significant statement. Scaling this result to the EUSO distance and acceptance, we obtain an expected threshold energy for the Cherenkov r/d signal of $8 \times 10^{19}$ eV. This estimate doesn't take into account the Rayleigh and Mie scattering and is linked to the particular location of the experiment, with shallow water close to the shore linked in an unknown way with the open ocean water. Nevertheless this is the first experimental detection of Cherenkov light from sea water, and demonstrates within the previous limits that the r/d Cherenkov light detection from Space is feasible with an energy threshold close to fluorescence.

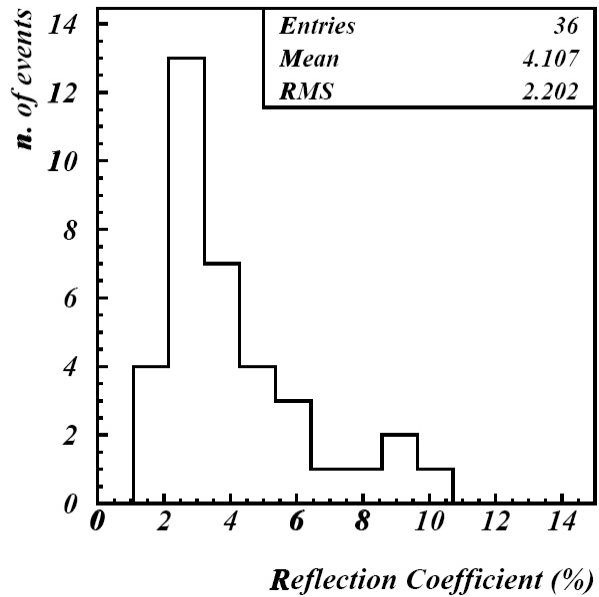

Figure 3: Distribution of the estimated r/d coefficient.

## Conclusions

The data sample collected during the ULTRA 2005 campaign is quite small, but the information collected for each event allows to be very confident on their genuineness. All the checks done on this sample after selection confirm this assumption. The problems arose with the direct Cherenkov light detector, forcing us to have Monte Carlo dependent results, this being the real limit of this test. On the basis of the experience gained during this campaign, we are confident that having the possibility of making another measurement campaign we could: a) improve the method of direct Cherenkov light measurement; b) increase the statistical significance of the present result; c) compare the r/d coefficient of different surfaces (deep water, desert surface, snow/ice, forest/grass) to cover the expected observational environment of a Space satellite.

## Acknowledgements

M. Dattoli would like to thank the National Institute for Astrophysics (INAF) for partly supporting her activity during this project.